\documentclass{iucrjournals}

\usepackage{amsmath}
\usepackage{booktabs,threeparttable,xcolor}
\usepackage{units}
\usepackage{placeins}

\nolinenumbers % arXiv does not accept line numbers (iucrjournals.cls enables them)

\title{\textit{lawless}: Normalization, Scaling, and Merging of Laue Data in the \textit{AIMLESS} Framework}

\author[a,b]{Aaron D. Finke\IUCrCemaillink{aaron.finke@ess.eu}\IUCrOrcidlink{0000-0002-6842-1122}}%

\affil[a]{European Spallation Source ERIC, Data Management and Scientific Computing, Kongens Lyngby, Denmark }
\affil[b]{European Spallation Source ERIC, Lund, Sweden }

\begin{document} 
\maketitle 

\begin{synopsis}
The new Laue normalization and scaling program \textit{lawless}, based on the classic \textit{AIMLESS} program, provides a convenient method for normalizing, scaling, and merging Laue data from both X-rays and neutrons, ensuring better map quality in particular for neutron Laue data by taking a softer hand to outlier rejection, compared to previous routines.
\end{synopsis}

\begin{abstract}
\textit{lawless}, a new program for wavelength normalization, scaling, and merging of Laue data from both X-rays and neutrons, is presented. Based on the \textit{AIMLESS} program, \textit{lawless} provides two methods of wavelength normalization via Chebyshev polynomials and Gaussian process regression, the latter of which is a more general method for various sources. Coupled with \textit{AIMLESS}'s gentler outlier rejection methodology and superior error modelling, \textit{lawless} provides better quality neutron maps than previous methods by focusing on completeness over better-looking merging statistics. Even as R-factors increase and CC1/2 numbers decrease, dataset completeness is the biggest limiting factor in neutron map quality and should be a primary focus in data reduction.
\end{abstract}


\section{Introduction}

Polychromatic diffraction in macromolecular crystallography (MX), which in the 1980s was the most promising method for collecting X-ray MX data\cite{helliwell_recording_1989}, has been largely, but not completely, obviated by the introduction of more powerful synchrotron light sources, the flux from whose monochromated beams are more than sufficient for the smallest crystal samples. Nonetheless, it retains utility for certain techniques. The "Laue" method uses a "white beam," which is a broad spectrum of X-ray or neutron radiation unconstrained by monochromation; the "quasi-Laue" method uses a constrained, "pink beam" of radiation of bandwidth ca. 5-30\%. (For reference, silicon monochromators at modern synchrotron MX beamlines provide bandwidths of ca. 0.01\%.) For polychromatic MX, quasi-Laue techniques with pink beams are preferred to simplify data processing and minimize spot overlap.

Quasi-Laue methods are the most popular technique for collecting neutron MX data. As of October 2026, there are 262 neutron MX entries in the PDB; of those, 186 are from Laue data, and 76 are from monochromatic data\cite{wwpdb_consortium_protein_2019}. Polychromatic methods are much better for neutron economy, a critical consideration given the relatively low flux of even the most powerful neutron sources. In addition, neutron time-of-flight (ToF) Laue methods can exploit the de Broglie matter-wave nature of neutrons to discriminate by wavelength, leading to "monochromatic images" from Laue data, with lower background and fewer reflection overlaps per image than true Laue data \cite{kono_current_2022}.

The development of new methods for indexing and integration of Laue data has experienced a Renaissance recently. The recently released \textit{laue-dials}\cite{hewitt_laue-dials_2024} is the first modern, general-purpose toolkit dedicated to Laue MX data, and one of the few since the Daresbury Laue software suite was introduced in 1995 \cite{campbell_lauegen_1995,campbell_lauegen_1998}. Advances in neutron Laue-ToF MX data processing has recently been led by the Mantid\cite{arnold_mantiddata_2014,coates_macromolecular_2020}, STARGazer,\cite{yano_application_2016,ohhara_development_2009} and DIALS \cite{McDonagh2026_DIALS} projects. The development of "pink-beam" serial crystallography, which uses bandpasses of ca. 5\%, has led to algorithmic improvements in indexing pink-beam data \cite{gevorkov_pinkindexer_2020}; the broader bandwidth of light used in pink-beam methods gives more Bragg reflections per exposure, which can improve crystal economy in serial methods. 

While indexing and integration methods have progressed, new methods in scaling and correction of Laue MX data has seen less progress. \textit{Laue-dials} does not use a scaling routine, and instead relies on a sister program, \textit{careless},\cite{dalton_unifying_2022} to do scaling and wavelength normalization. \textit{Careless} relies on machine learning methods and is quite different from the typical scaling methods, relying on neural networks trained on Wilson-statistical priors. Most neutron Laue datasets submitted to the PDB still use the \textit{LSCALE} \cite{arzt_lscale_1999} or \textit{LAUENORM} routines in the Daresbury Laue suite \cite{campbell_lauegen_1998}, coupled with the \textit{SCALA} \cite{evans_scaling_2006} program for scaling and merging. This is suboptimal for a few reasons. First, these programs handle weak data improperly, often throwing away reflections below a certain $I/\sigma(I)$ threshold. Second, it relies too heavily on R-factor statistics to define resolution cutoffs, instead of using clustering analysis which has been shown to better distinguish real data from noise in Bragg reflection data. Much has been learned about the "correct" treatment of weak reflection data in the past 15 years, but these lessons appear to have not gained traction in the neutron MX sphere \cite{diederichs_better_2013,karplus_assessing_2015,karplus_linking_2012}.

These lessons also apply to neutron Laue MX. Here, \textit{lawless}\footnote{"lawless" is simply a portmanteau of "Laue-\textit{AIMLESS}." The program \textit{lawless} is styled in all-lowercase, for no reason aside from aesthetics.} is presented-- an improved program for wavelength normalization, scaling, and merging of Laue crystallographic data. It is built on top of the \textit{AIMLESS} program distributed with the \textit{CCP4} suite\cite{agirre_ccp4_2023}, which is its primary application for scaling and merging monochromatic MX data. \textit{lawless} provides wavelength normalization with two different methods: Chebyshev polynomial fitting, and Gaussian process regression. In addition, it provides a few additional wavelength-dependent scale factors for error modelling. The normalization and scaling workflow in \textit{lawless} can provide better results than previous methods, especially for weak neutron Laue MX data, based on the quality of nuclear density maps generated from the merged data. 

\section{The \textit{lawless} program}

\subsection{Requirements}

A scaling and merging routine for Laue MX data, especially for neutron Laue MX data, must satisfy a few requirements:

\begin{itemize}
    \item Handle wavelength normalization from various X-ray and neutron sources of arbitrary bandwidths;
    \item Perform all corrections needed from integrated intensities and directly output data for use in structural refinement;
    \item Prioritize completeness over multiplicity/merging statistics like $R_{merge}$ or $R_{meas}$;
    \item Provide modern merging statistical measures such as CC\nicefrac{1}{2} to help determine resolution cutoffs;
    \item Have a flexible, refinable error model to handle the sometimes large uncertainties inherent in neutron Laue MX measurements.
\end{itemize}

Fortunately, the \textit{AIMLESS} program \cite{evans_how_2013} satisfies the latter four criteria already. \textit{AIMLESS} is a modernized version of the classic \textit{SCALA} program\cite{evans_scaling_2006}, both written by Phil Evans, for scaling and merging MX data. \textit{AIMLESS} was written in C++ and is designed to be modular, simplifying incorporation of new routines-- so building a Laue normalization routine on top of \textit{AIMLESS} is very much in line with the original program's ethos. The scaling model, outlier rejection, error model, and statistical calculations in \textit{AIMLESS} remain the current standard for MX data reduction. The methods in \textit{AIMLESS} are flexible enough to accommodate non-rotation "batch" data, such as serial crystallography. Indeed, they also work well for Laue data, once the data have been properly corrected.

The author refers readers to the original \textit{AIMLESS} paper for a complete description of the scaling, outlier, and error models the program uses; \cite{evans_how_2013} unless otherwise stated, they are used unchanged in \textit{lawless}. 

\subsection{\textit{lawless} as a modified version of \textit{AIMLESS}}

\textit{lawless} sits on top of the \textit{AIMLESS} code base and is best described as a modification of the latter program. Writing of the new code was assisted by Claude Opus 5.5 (Anthropic; model identifier claude-opus-5-5), used through Claude Code, and reviewed, revised, and tested by the author. Compared to compilation requirements for \textit{AIMLESS}, \textit{lawless} only requires one additional library: the \textit{Eigen} library \cite{eigenweb}, for the Gaussian process regression normalization (see Section \ref{GPR}). \textit{CCP4} users already have all the required dependencies: \textit{Eigen} is distributed with \textit{CCP4}. \textit{lawless} is distributed as free and open-source software with the BSD 3-clause licence. The source code archive, as well as the repository link, can be found at https://doi.org/10.5281/zenodo.23929955. 

\textit{lawless} runs identically to \textit{AIMLESS} v0.8.6 (release version in CCP4 v9 \cite{agirre_ccp4_2023}) in the absence of Laue-specific instructions, so it can be used as a drop-in replacement for \textit{AIMLESS} if one wishes to use it for both monochromatic and Laue data. 

In this text, \textit{lawless} instructions are given in all caps in teletype font, e.g. \texttt{LAUE NORMGPR}. 

\subsection{Extended scale model}

Per-reflection scale factors $g$ are refined in \textit{AIMLESS} as a BFGS optimization of several parameters. \textit{lawless} adds a wavelength-dependent term to the scaling model:

\begin{equation}
    g = ps \times bs \times ss \times ds \times \boldsymbol{ws(\lambda)}
    \label{eq:scale_factor}
\end{equation}

The first four terms in $g$ are identical to those in \textit{AIMLESS}, with one modification. $ps$ is the primary multiplicative scale factor based on rotation (or batch parameter); $bs$ is the isotropic B-factor via $\exp(-B\cdot s^{2})$; $ss$ is the secondary beam correction, which has been modified to reflect the polychromatic input data (see \ref{secondarybeam}); $ds$ is detector gain correction. $ws(\lambda)$ is the new wavelength-dependent correction factor introduced here. For monochromatic data, $ws(\lambda)=1$. 

\subsection{Wavelength normalization}

The intensity \textit{I} of a Bragg reflection of a stationary crystal from a polychromatic beam is defined by Eq. \ref{eq:polychromiatic_intensity} \cite{buras_relations_1975}:

\begin{equation}
    I = i_{0}(\lambda) V N^{2} \lvert F \rvert^{2} p \left( \frac{\lambda^{4}}{2 \sin^{2}(\theta)} \right)
    \label{eq:polychromiatic_intensity}
\end{equation}

where $i_{0}(\lambda)$ is the intensity of the incoming beam as a function of wavelength $\lambda$, $V$ is the crystal volume, $N$ is the number of unit cells per crystal volume, $F$ is the structure factor, and $p$ is the polarization factor ($p=1$ for neutrons). The $\lambda^4/2\sin^2(\theta)$ term is the Lorentz correction factor \cite{iizumi_lorentz_1973,michels-clark_expanding_2016}. Two of these terms are wavelength-dependent, along with the structure factor: the quartic power relationship and the intensity spectrum. The two wavelength normalization methods in \textit{lawless} can absorb both of these corrections simultaneously. In addition, wavelength-dependent parameters such as detector response, measured empirically for neutron diffractometers by spectroscopic measurements of vanadium metal, can also be absorbed by the wavelength correction. The correction factor is then normalized to a reference wavelength-- usually the spectral maximum-- to give $ws$.

\textit{lawless} has two modes for wavelength normalization.

\subsubsection{Chebyshev polynomial normalization}

\textit{lawless} provides Chebyshev polynomial normalization (\texttt{LAUE NORMCHEB}) with an arbitrary number of coefficients and five non-overlapping wavelength ranges. A wavelength-only pre-pass is done to decorrelate $ws(\lambda)$ from the per-batch scales before joint refinement. The polynomials are then refined jointly with other scaling parameters in the BFGS step (eq. \ref{eq:scale_factor}). \textit{lawless} keeps reflections outside the given wavelength ranges and sets $ws=1$ for them. 

The Chebyshev implementation in \textit{lawless} is slightly different compared to previous Laue scaling programs. \textit{LAUENORM} refines discrete scale factors for wavelength bins from overlaps between symmetry equivalents, with one bin fixed at unity, and least-squares fits these with an ordinary power series $A(\lambda)=\sum_k p_k\lambda^k$ for each wavelength range. \textit{LSCALE} models the wavelength normalization as a Chebyshev series in linear space, $A(\lambda) = \tfrac{1}{2}c_0 + \sum_{k\ge1} c_k T_k(z)$, where $z$ maps the wavelength range onto $[-1,1]$. It is applied as $I_{sc} = SC \cdot I$ with $SC = A(\lambda)/A(\lambda_{ref})$. Nothing in these forms keep $A(\lambda)$ positive; where data are sparse, typically at the edges of the bandpass, the fitted polynomial can approach zero or go negative, which gives meaningless scaled intensities and makes the refinement unstable.

\textit{lawless} instead fits the \textit{logarithm} of the wavelength response as a Chebyshev series, $h(\lambda) = \sum_k a_k T_k(z)$, and applies $ws = \exp(h(\lambda) - h(\lambda_{ref}))$ as a divisor of the intensity, so $ws$ is effectively $1/SC$. The exponential guarantees $ws > 0$ for any coefficient values. By itself, Chebyshev normalization in log space does not stop the curve from diverging in sparse-data regions, but it is nonetheless a reasonable guard. 

\subsubsection{Gaussian Process Regression}\label{GPR}

Gaussian process regression (GPR) is a non-parametric Bayesian method that places a prior directly over functions rather than assuming a fixed functional form \cite{rasmussen_gaussian_2008}. A kernel sets how strongly values at nearby wavelengths are correlated, and conditioning on the observations gives a posterior mean and uncertainty at every wavelength. The kernel hyperparameters are chosen by maximizing the marginal likelihood, so the flexibility of the curve is set by the data rather than by a polynomial degree. This suits Laue wavelength normalization, particularly for broadband or white-beam applications. A polynomial of fixed degree must trade fidelity in the steep, densely sampled part of the spectrum against stability in the sparse tails, where it tends to diverge. By contrast, in a GP the influence of each observation is local, and away from the data the curve relaxes toward its prior mean. The posterior uncertainty also widens where observations are few, showing where the curve is poorly determined.

In \textit{lawless} GPR normalization (\texttt{LAUE NORMGPR}) the GP is fitted in a pre-pass to binned leave-one-out intensity ratios $\rho=I_{obs}/\langle I_{mates} \rangle$ and then enters the scale factor $g$ as a fixed term. Unlike Chebyshev coefficients, the GP is not refined during scaling. The ratios are grouped into bins of equal count, with a cap on bin width in $\log\lambda$. As in the Chebyshev model, the fit is to the logarithm of the response, which guarantees a positive correction. The GP has a constant mean and a squared-exponential (default) or Matérn-3/2 covariance function. The length scale, signal variance, and a noise-inflation factor are chosen jointly by maximising the log marginal likelihood over a grid, each candidate being evaluated by Cholesky factorisation of the covariance matrix. The correction is normalized to a reference wavelength, $ws(\lambda) = \exp(g(\lambda) - g(\lambda_{ref}))$; if no reference is given, $g(\lambda_{ref})$ is replaced by the constant GP mean, so that $ws = 1$ at the average response. Reflections outside the fitted wavelength range receive $ws = 1$, but by default, the entire wavelength range is included. A detailed description of the GPR in \textit{lawless} is provided in Appendix \ref{GPR_maths}.

There is also a \texttt{LAUE NORMGPRPERRUN} mode that does a per-run residual curve fitting $ws(\lambda, run) = ws_{global} \times ws_{run}$. This may be useful to see divergences in wavelength fitting curves per exposure, from e.g. optics vibrations or neutron chopper phase errors. 

\subsection{Wavelength-aware secondary beam geometry}\label{secondarybeam}

Secondary beam corrections $\mathbf{s}_2$ in \textit{AIMLESS} depend on knowing the correct wavelength, usually given by the batch wavelength in the input mtz file. For Laue data this is obviously incorrect as each observation has its own wavelength; in \textit{lawless} \texttt{LAUE} mode, the observed wavelength for each reflection is passed instead.

This correction is necessary because the secondary beam direction cosines drive the absorption and surface corrections. For example, in the case of a Laue measurement of wavelength range $\lambda=$ 2.10–3.90 \AA (ie. the LADI-III instrument), if the median wavelength is used, the correction is evaluated in the wrong direction by a median of $6.4^\circ$ and up to $18^\circ$. 

\subsection{Wavelength term in the SD correction}

The standard error estimates of each intensity observation $\sigma'(I_{hl})$ are calculated in \textit{AIMLESS} by

\begin{equation}
    \sigma'(I_{hl}) = \mathrm{SdFac}\left[\sigma(I_{hl})^2 + \mathrm{SdB},\langle I_h\rangle + (\mathrm{SdAdd},\langle I_h\rangle)^2\right]^{1/2}
    \label{eq:error_estimate}
\end{equation}

Variance trends that are wavelength-dependent cannot be fixed by the scaling model alone. This can include wavelength normalization (recall that GPR gives an uncertainty), absorption, and extinction effects. In addition, wavelength-dependent integration quality is not represented in $\sigma(I)$ and Eq. \ref{eq:error_estimate} cannot absorb them. An optional per-run power law in $\lambda$ can test for such residuals. 

For Laue data, the corrected error estimate of Eq. \ref{eq:error_estimate} can be further multiplied by $\left(\frac{\lambda_{hl}}{\lambda_{\mathrm{ref}}}\right)^{\mathrm{SdLam}}$ with one exponent per run, fitted after SdFac, SdB, and SdAdd from the slope of $\log\chi^2$ against $\log(\lambda/\lambda_{\mathrm{ref}})$ in wavelength bins. This wavelength-dependent term is optional and supplied with the \texttt{SDLAMBDA} instruction alongside \texttt{SDCORRECTION}.

\subsection{Wavelength normalization without scaling}

\textit{lawless} has a \texttt{LAMBDAONLY} mode where the data are corrected only by the initial wavelength-only pre-pass, and the corrected data can be output as either merged or unmerged data. In \texttt{LAMBDAONLY} mode with GPR, only the wavelength normalization is applied, without other scale factors or outlier rejection. Because the GP is fitted to unscaled intensities and held constant, the wavelength correction $ws(\lambda)$ is identical with or without subsequent scaling. This differs from the Chebyshev model, whose coefficients are refined further in the main BFGS step. 

\section{Test Data}

\textit{lawless} was tested on a variety of X-ray and neutron Laue MX datasets; the results of the optimized runs are in Table \ref{tab:data}. The DHFR\cite{hewitt_x-ray_DHFR} and HEWL\cite{hewitt_x-ray_HEWL} datasets are X-ray Laue datasets from the BioCARS beamline at APS \cite{henning_biocars_2024}. CA-thio and xylanase\cite{wan_direct_2015} are from the (now-defunct) PCS neutron Laue-TOF instrument at LANSCE\cite{langan_pcs_2004}; two, dTIM-2PG and dTIM-PGH \cite{kelpsas_perdeuteration_2019}, are from the LADI-III quasi-Laue neutron diffractometer at ILL\cite{blakeley_neutron_2010}; and dMPro-KB5 \cite{bhandari_noncovalent_2026} is from the MaNDi Laue-TOF neutron diffractometer at SNS\cite{coates_macromolecular_2020}. 

\begin{table*}[htbp]
\caption{Data collection and merging statistics for the test data sets.}
\label{tab:data}
\begin{threeparttable}
\footnotesize
\setlength{\tabcolsep}{1.7pt}
\begin{tabular}{lccccccc}
\toprule
 & DHFR\tnote{a} & HEWL\tnote{b} & CA--thio\tnote{c} & Xylanase\tnote{d} & dTIM--2PG\tnote{e} & dTIM--PGH\tnote{e} & dMPro--KB5\tnote{f} \\
\midrule
\multicolumn{8}{l}{\textit{Data collection}} \\
Protein & \begin{tabular}[t]{@{}c@{}}DHFR--NADP\textsuperscript{+}\\--folate\end{tabular} & HEWL & \begin{tabular}[t]{@{}c@{}}hCA II--\\thiophene\end{tabular} & \begin{tabular}[t]{@{}c@{}}XynII\\N44D\end{tabular} & TIM--2PG & TIM--PGH & \begin{tabular}[t]{@{}c@{}}M\textsuperscript{pro}--KB5\end{tabular} \\
Instrument & \begin{tabular}[t]{@{}c@{}}BioCARS,\\APS\end{tabular} & \begin{tabular}[t]{@{}c@{}}BioCARS,\\APS\end{tabular} & \begin{tabular}[t]{@{}c@{}}PCS,\\LANSCE\end{tabular} & \begin{tabular}[t]{@{}c@{}}PCS,\\LANSCE\end{tabular} & \begin{tabular}[t]{@{}c@{}}LADI-III,\\ILL\end{tabular} & \begin{tabular}[t]{@{}c@{}}LADI-III,\\ILL\end{tabular} & \begin{tabular}[t]{@{}c@{}}MaNDi,\\SNS\end{tabular} \\
Wavelength (\AA) & 0.96--1.15 & 0.95--1.30 & 0.73--6.20 & 0.73--6.38 & 2.54--3.60 & 2.70--3.60 & 2.10--3.90 \\
Images (batches) & 175 & 371 & 40 & 24 & 49 & 41 & 15 \\
Space group & $P2_12_12_1$ & $P4_32_12$ & $P2_1$ & $P2_12_12_1$ & $C2$ & $C2$ & $I2$ \\
$a$, $b$, $c$ (\AA) & 34.3,\,45.5,\,99.0 & 79.5,\,79.5,\,37.8 & 42.9,\,41.6,\,72.9 & 49.6,\,60.0,\,70.6 & 98.3,\,52.4,\,58.4 & 98.3,\,52.7,\,58.7 & 55.2,\,81.4,\,88.6 \\
$\alpha$, $\beta$, $\gamma$ ($^\circ$) & 90,\,90.0,\,90 & 90,\,90.0,\,90 & 90,\,104.5,\,90 & 90,\,90.0,\,90 & 90,\,118.0,\,90 & 90,\,117.9,\,90 & 90,\,96.6,\,90 \\
Data Reduction & \textit{laue-dials} & \textit{laue-dials} & \textit{d*TREK}\tnote{g} & \textit{d*TREK}\tnote{g} & \textit{LAUEGEN} & \textit{LAUEGEN} & \textit{Mantid} \\
\addlinespace
\multicolumn{8}{l}{\textit{Scaling and merging (\textit{lawless} 0.1.6)}} \\
$\lambda$ normalization & GPR & GPR & GPR & GPR & GPR & GPR & GPR \\
Resolution (\AA) & 49.5--1.71 & 35.6--1.79 & 25.6--2.00 & 19.2--2.40 & 30.4--1.75 & 30.6--1.80 & 13.9--2.30 \\
Outer shell (\AA) & 1.74--1.71 & 1.83--1.79 & 2.05--2.00 & 2.49--2.40 & 1.78--1.75 & 1.84--1.80 & 2.38--2.30 \\
Observations & 198\,114 & 1\,083\,974 & 72\,444 & 47\,375 & 327\,833 & 217\,344 & 61\,809 \\
Unique reflections & 17\,110 & 11\,836 & 15\,192 & 8\,004 & 24\,785 & 22\,222 & 15\,940 \\
Multiplicity & 11.6 (2.6) & 91.6 (21.5) & 4.8 (3.9) & 5.9 (4.8) & 13.2 (15.1) & 9.8 (13.4) & 3.9 (4.0) \\
Completeness (\%) & 99.0 (84.4) & 99.8 (98.4) & 89.9 (83.3) & 93.6 (91.9) & 93.3 (83.6) & 90.5 (81.8) & 91.7 (83.9) \\
$\langle I/\sigma(I)\rangle$ & 7.1 (1.7) & 22.4 (5.0) & 3.6 (1.5) & 3.6 (1.2) & 4.5 (1.7) & 2.1 (0.5) & 5.4 (3.9) \\
CC$_{1/2}$ & 0.974 (0.760) & 0.988 (0.963) & 0.926 (0.289) & 0.883 (0.323) & 0.936 (0.754) & 0.826 (0.178) & 0.970 (0.776) \\
$R_\mathrm{meas}$ & 0.283 (0.563) & 0.196 (0.447) & 0.392 (1.005) & 0.462 (1.488) & 0.725 (2.493) & 1.449 (7.390) & 0.175 (0.310) \\
$R_\mathrm{pim}$ & 0.081 (0.306) & 0.022 (0.091) & 0.161 (0.451) & 0.180 (0.621) & 0.190 (0.615) & 0.410 (1.772) & 0.080 (0.135) \\
$\langle\chi^2\rangle$ & 0.81 (0.77) & 1.04 (0.85) & 0.89 (0.74) & 0.87 (0.81) & 0.91 (0.95) & 0.93 (0.90) & 0.75 (0.65) \\
$d_\mathrm{min}$, CC$_{1/2}{=}0.3$ (\AA) & 1.71 & 1.79 & 2.04 & 2.40 & 1.75 & 1.98 & 2.30 \\
SdFac, SdB, SdAdd & 0.77,\,-0.56,\,0.38 & 0.41,\,1.11,\,0.51 & 2.61,\,0.16,\,0.10 & 3.32,\,0.24,\,0.09 & 1.34,\,1.62,\,0.23 & 1.27,\,2.36,\,0.31 & 0.82,\,0.09,\,0.29 \\
Outliers rejected & 21\,110 & 6\,898 & 490 & 350 & 1\,722 & 512 & 893 \\
\bottomrule
\end{tabular}
\begin{tablenotes}
\footnotesize
\item All data sets were scaled and merged with \textit{lawless} 0.1.6. Values in parentheses are for the outermost resolution shell. Statistics are with merged Friedel mates. The error model for all data sets is refined (\texttt{SDCORRECTION REFINE}).
\item[a] DHFR from \cite{hewitt_x-ray_DHFR}: images 1--4 of 179 are excluded (their scales diverge in every per-image scale model).
\item[b] HEWL from \cite{hewitt_x-ray_HEWL}: one pass of 375 images 1$^\circ$ apart; 3 failed polychromatic refinement and the last (short exposure) is excluded. Smooth scales along the rotation (\texttt{SCALES ROTATION SPACING 5 BFACTOR ON}); \texttt{REJECT 4 LARGER}; \texttt{ANOMALOUS ON}.
\item[c] hCAII with thiophene inhibitor from S. Zoë Fisher, personal communication.
\item[d] Endo-1,4-$\beta$-xylanase II (XynII) N44D from \cite{wan_direct_2015}: the 24 \textit{d*TREK} \texttt{.ref} files, negative intensities kept.
\item[e] Perdeuterated TIM with 2PG and PGH bound from \cite{kelpsas_perdeuteration_2019}.
\item[f] Perdeuterated MPro-KB5 from \cite{bhandari_noncovalent_2026}.
\item[g] \textit{d*TREK} as adapted for the PCS wavelength-resolved Laue data \cite{langan_protein_2004}. Unfortunately, the source binaries for these programs have been lost. 
\end{tablenotes}
\end{threeparttable}
\end{table*}

In general, \textit{lawless} gave normalized, scaled, and merged datasets with good merging statistics, with a significant improvement in data completeness due to a gentler outlier removal method. The merging statistics do not \textit{look} better than the published data in many cases (See Table \ref{tab:legacy}), especially with the neutron Laue datasets. However, as we will see, merging statistics are not always a reliable indicator of data quality or utility.

\begin{table*}[htbp]
\caption{\textit{lawless} compared with the established pipeline for each neutron Laue data set.}
\label{tab:legacy}
\begin{threeparttable}
\footnotesize
\setlength{\tabcolsep}{4pt}
\begin{tabular}{lrrrrrrc}
\toprule
 & Obs. & Unique & Compl. (\%) & Mult. & $\langle I/\sigma\rangle$ & $R_\mathrm{pim}$ & CC$_{1/2}$ (outer) \\
\midrule
\multicolumn{8}{l}{\textbf{dTIM--2PG}} \\
\textit{LSCALE}+\textit{SCALA}\tnote{a} & 128\,793 & 21\,821 & 82.8 & 5.9 & 9.2 & 0.052 & 0.987 (0.978) \\
\textit{lawless} & 327\,531 & 24\,718 & 93.6 & 13.3 & 4.5 & 0.190 & 0.939 (0.740) \\
\textit{lawless}, matched\tnote{b} & 290\,870 & 21\,814 & 82.7 & 13.3 & 5.0 & 0.170 & 0.947 (0.827) \\
\addlinespace
\multicolumn{8}{l}{\textbf{dTIM--PGH}} \\
\textit{LSCALE}+\textit{SCALA}, published\tnote{c} & 99\,160 & 19\,145 & 78.0 & 5.2 & 3.8 & 0.118 & 0.941 (0.638) \\
\textit{LSCALE}+\textit{SCALA}, CUTOFF 0.5 & 53\,003 & 16\,076 & 65.5 & 3.3 & 6.5 & 0.078 & 0.961 (0.884) \\
\textit{lawless} & 217\,344 & 22\,222 & 90.5 & 9.8 & 2.1 & 0.410 & 0.829 (0.195) \\
\textit{lawless}, matched\tnote{b} & 191\,426 & 19\,141 & 78.0 & 10.0 & 2.5 & 0.351 & 0.843 (0.311) \\
\addlinespace
\multicolumn{8}{l}{\textbf{dMPro--KB5} }\\
\textit{Mantid+LAUENORM} & 62\,496 & 15\,943 & 91.9 & 3.9 & 6.5 & 0.091 & 0.952 (0.743) \\
\textit{lawless} & 61\,809 & 15\,940 & 91.9 & 3.9 & 5.1 & 0.079 & 0.970 (0.753) \\
\addlinespace
\multicolumn{8}{l}{\textbf{Xylanase}} \\
\textit{LAUENORM}+\textit{SCALA}\tnote{d} & 41\,946 & 12\,101 & 83.7 & 3.5 & 5.2 & 0.124 & 0.845 (0.137) \\
\textit{lawless} & 71\,206 & 13\,437 & 92.4 & 5.3 & 2.5 & 0.272 & 0.806 (0.119) \\
\textit{lawless}, matched\tnote{b} & 68\,445 & 12\,100 & 83.7 & 5.7 & 2.8 & 0.262 & 0.806 ($-$0.029) \\
\textit{lawless}, negatives set to 0\tnote{e} & 71\,167 & 13\,436 & 92.4 & 5.3 & 3.0 & 0.203 & 0.868 (0.105) \\
\bottomrule
\end{tabular}
\begin{tablenotes}
\footnotesize
\item[a] Archived \textit{LSCALE} output (CUTOFF 0.5) of the same 49 \textit{LAUEGEN} images, merged by the original \textit{SCALA} script.
\item[b] \textit{lawless} observations restricted to the unique reflections present in the \textit{LSCALE} (xylanase: \textit{SCALA}) output, so that completeness is matched.
\item[c] \textit{LSCALE} CUTOFF 2 + \textit{SCALA} reject 3; reproduces the published neutron data \cite{kelpsas_perdeuteration_2019} to CC 0.99996.
\item[d] The PCS standard processing: the archived \textit{LAUENORM} output of the same 24 \textit{d*TREK} frames (negative intensities set to 0), merged by the archived \textit{SCALA} script (batch scales, reject 2) rerun with unmerged output; 17.9--2.0~\AA. \textit{lawless}: the Table~\ref{tab:data} deck to 2.0~\AA.
\item[e] \textit{lawless}, same deck, on the \textit{d*TREK} intensities that \textit{LAUENORM} read: every negative intensity (24\% of the observations) set to 0.
\end{tablenotes}
\end{threeparttable}
\end{table*}

\section{Results and Discussion}

\subsection{Monochromatic regression test}

As stated above, running \textit{lawless} without the \texttt{LAUE} keyword should behave identically to \textit{AIMLESS}. A monochromatic rotation dataset of thaumatin (BioMAX, MAX IV Laboratory) was used as a regression test to ensure that the new features of \textit{lawless} do not affect the standard, monochromatic scaling routine in \textit{AIMLESS} (Table \ref{tab:thaumatin}). Results were identical to \textit{AIMLESS} v0.8.2 from the CCP4 v9 release \cite{agirre_ccp4_2023}, well within rounding errors (Table \ref{tab:thaumatin}).

\subsection{Wavelength normalization: GPR vs Chebyshev}

\begin{figure}[htbp]
    \centering
    \includegraphics[width=1\linewidth]{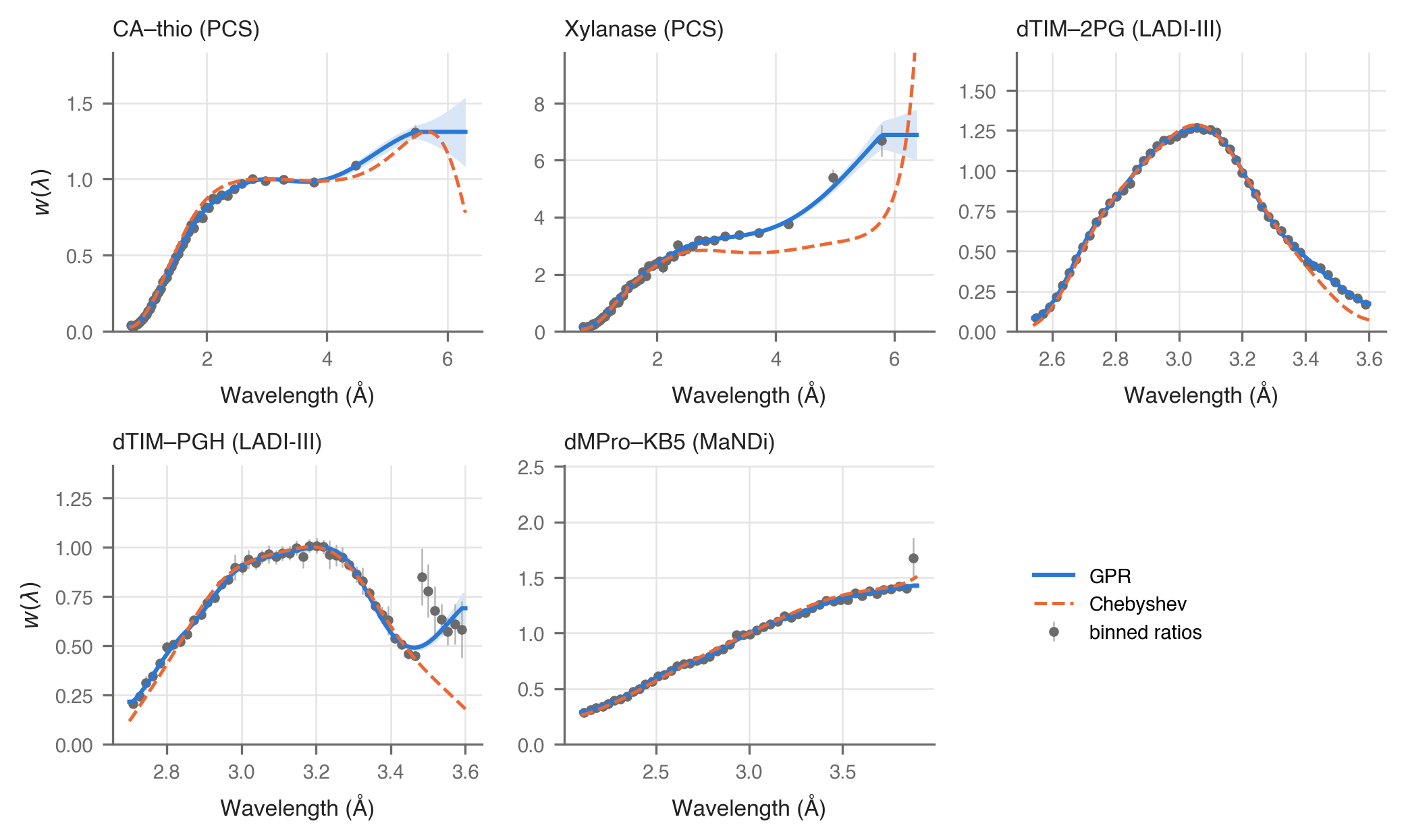}
    \caption{Wavelength normalization curves $w(\lambda)$ for each neutron Laue data set, fitted by Chebyshev polynomials (orange, dashed) and GPR (blue, with the 1$\sigma$ posterior band) with the binned leave-one-out intensity ratios the GP is fitted to the grey dots ($\pm1\sigma$).}
    \label{fig:lambda_norm}
\end{figure}

Wavelength normalization using GPR and Chebyshev methods was done on all datasets (Figure \ref{fig:lambda_norm}). In terms of computation time, the two methods are roughly equal: while the GPR pre-pass is slower than the Chebyshev pre-pass, wavelength normalization in GPR does not add parameters to the BFGS optimization like Chebyshev does, so the total compute time evens out. Wavelength normalization takes significantly less time than scaling in \textit{lawless}. 

In the cases where the X-ray/neutron spectra are smooth and easily fit to a polynomial, there is little difference between the Chebyshev and GPR normalization results. This is the case for the data from LADI-III and MaNDi. This is not always the case, however (Table \ref{tab:gprcheb}). 

\begin{table*}[htbp]
\caption{Merging statistics with Gaussian-process and Chebyshev wavelength normalization.}
\label{tab:gprcheb}
\begin{threeparttable}
\footnotesize
\setlength{\tabcolsep}{4pt}
\begin{tabular}{llcrcccc}
\toprule
Data set & Method & Degree\tnote{a} & Obs. & $\langle I/\sigma\rangle$ & $R_\mathrm{pim}$ & CC$_{1/2}$ (outer) & $\langle\chi^2\rangle$ \\
\midrule
DHFR & GPR & -- & 198\,114 & 7.1 & 0.081 & 0.974 (0.760) & 0.81 \\
 & Chebyshev & 9 & 197\,936 & 5.3 & 0.154 & 0.884 (0.758) & 0.84 \\
\addlinespace
HEWL & GPR & -- & 1\,083\,974 & 22.4 & 0.022 & 0.988 (0.963) & 1.04 \\
 & Chebyshev & 9 & -- & -- & -- & -- & -- \\
\addlinespace
CA--thio & GPR & -- & 72\,444 & 3.6 & 0.161 & 0.926 (0.289) & 0.89 \\
 & Chebyshev & 2 $\times$ 6 & 72\,408 & 3.6 & 0.163 & 0.929 (0.281) & 0.87 \\
\addlinespace
Xylanase & GPR & -- & 47\,375 & 3.6 & 0.180 & 0.883 (0.323) & 0.87 \\
 & Chebyshev & 2 $\times$ 6 & 42\,381 & 11.0 & 0.201 & 0.878 (0.374) & 1.23 \\
\addlinespace
dTIM--2PG & GPR & -- & 327\,833 & 4.5 & 0.190 & 0.936 (0.754) & 0.91 \\
 & Chebyshev & 9 & 327\,072 & 4.6 & 0.232 & 0.933 (0.795) & 0.90 \\
\addlinespace
dTIM--PGH & GPR & -- & 217\,344 & 2.1 & 0.410 & 0.826 (0.178) & 0.93 \\
 & Chebyshev & 9 & 217\,129 & 2.2 & 0.465 & 0.808 (0.166) & 0.92 \\
\addlinespace
dMPro--KB5 & GPR & -- & 61\,809 & 5.4 & 0.080 & 0.970 (0.776) & 0.75 \\
 & Chebyshev & 6 & 61\,777 & 5.4 & 0.079 & 0.973 (0.746) & 0.74 \\
\bottomrule
\end{tabular}
\begin{tablenotes}
\footnotesize
\item[a] Chebyshev polynomial degree; $2\times6$: two wavelength ranges split at 1.5~\AA. For HEWL, degree 6 fails (the polynomial approaches zero at the band edges). For xylanase, the error model refined with Chebyshev normalization has a negative SdAdd, which inflates $\langle I/\sigma\rangle$.
\end{tablenotes}
\end{threeparttable}
\end{table*}

The CA-thio and xylanase datasets from PCS require two Chebyshev functions to be properly modelled. This is because of PCS's unique two-chopper system, which separately provided higher-energy, thermal neutrons ($\lambda<1.5\AA$) and lower-energy, cold neutrons ($\lambda>1.5\AA$), with an abrupt changeover at 1.5 \AA \cite{langan_pcs_2004}. While each individual chopper spectrum was smooth, the combined spectrum was not. By contrast, GPR, being non-parametric, can handle the relative lack of smoothness in the PCS curve and only one GPR function needed to be modelled for the entire spectrum. In addition, the sparse data at high wavelength leads Chebyshev fits to diverge, while in GPR, it merely increases the $\sigma$ of the region fit.

\subsection{Refining the GP normalization routine}

\begin{figure}[htbp]
    \centering
    \includegraphics[width=0.75\linewidth]{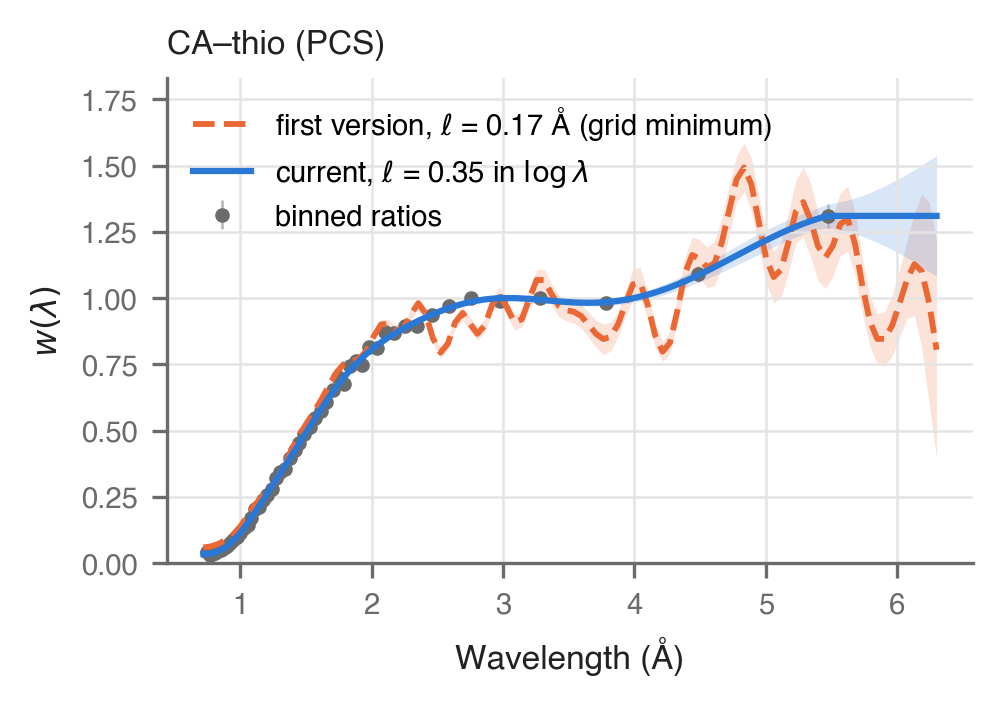}
    \caption{GPR wavelength normalization $w(\lambda)$ for CA–thio (PCS), first implementation vs the current one (lawless v0.1.6), each with its $\pm1\sigma$ posterior band, and the binned leave-one-out intensity ratios of the current fit (grey, $\pm1\sigma$).}
    \label{fig:gp_oscillation}
\end{figure}

GPR is a generic method that handles noisy, weak data differently than other regression methods. Tweaking the GP parameters is necessary to make a physically sensible model. The initial attempt at GP normalization and its optimized result for CA-thio is shown in Fig. \ref{fig:gp_oscillation}. The initial GPR result was a highly oscillating $ws(\lambda)$ which bore little physical resemblance to the true spectrum. The cause boiled down to a few issues, which were fixed by debiasing the bins from weaker observations (see Appendix \ref{GPR_maths}). In addition, evaluating the kernel length scale in $\ln(\lambda)$ rather than linearly in $\lambda$ helped model the sharp changes in the spectrum at shorter wavelength, in the case of the PCS and BioCARS data. 

\subsection{Pink beam X-ray Laue datasets: treatment of anomalous signal}\label{pinkbeam}

\begin{figure}[htbp]
    \centering
    \includegraphics[width=0.5\linewidth]{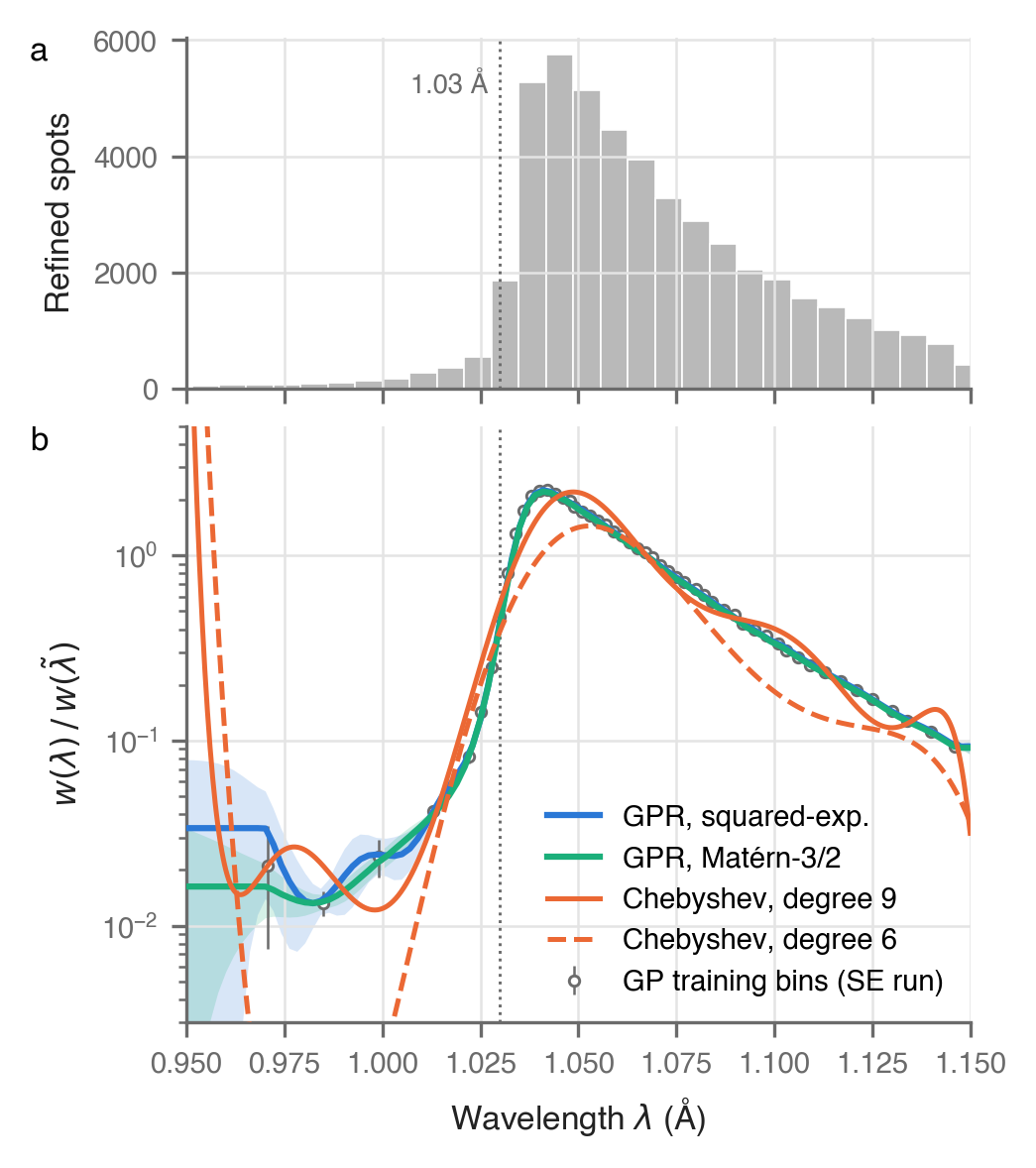}
    \caption{a) The spectrum of BioCARS based on histogram of number of refined spots of DHFR, from \textit{laue-dials}. b) Attempts to fit the BioCARS spectrum with Chebyshev polynomials and GPR fitting of squared-exponential and Mat\'ern-3/2 kernels. Posterior bands of GPR fits and error bars for training bins (open circles) are $\pm1\sigma$.}
    \label{fig:dhfr_spectrum}
\end{figure}

The intensity of the X-ray spectrum at BioCARS rises about two orders of magnitude between 1.2-1.4 \AA, making it difficult to model with Chebyshev polynomials of degree $<$ 9 (Fig. \ref{fig:dhfr_spectrum}). Scaling the DHFR dataset proved difficult as well because of this abrupt increase of intensity. Reflections measured at wavelengths $<1.04\ \mathrm{\AA}$ had significantly lower intensities than equivalent reflections recorded at higher wavelengths, leading to 10-25$\times$ disagreement with their partners. In \textit{LSCALE}, this problem is handled by simply removing the weak reflections; for DHFR, 11\% of reflections were rejected by \textit{LSCALE} outright, including the problematic ones, and Chebyshev normalization and scaling proceeded smoothly. In \textit{AIMLESS} and \textit{lawless}, however, the $\sigma$-based outlier removal cannot tell which observation is "wrong", and if a reflection is measured twice, by default it keeps both of a disagreeing pair. The batch scales then fit both reflections, dropping the low-resolution CC\nicefrac{1}{2} to 0.38–0.64 regardless of method. Rejecting one of the problematic pair (with \texttt{REJECT LARGER} or \texttt{REJECT SMALLER}) fixes this issue without having to do a blanket rejection of weak reflections, removing only a few hundred of the 198 000 reflections in the DHFR dataset. Using \texttt{REJECT LARGER} gave slightly better merging statistics than \texttt{REJECT SMALLER}.

The HEWL "serial" dataset from BioCARS fares similarly, though with far more reflections to be scaled. The HEWL dataset has far too many reflections to scale with \textit{LSCALE}; even when the program was modified to handle additional reflections, least-squares refinement crashed. Thus, the author decided to compare the \textit{lawless} result to the published \textit{careless} output \cite{dalton_unifying_2022}. Since the HEWL dataset contains anomalous signal from the sulfur atoms, the anomalous maps from the refined structure were investigated (Table \ref{tab:hewl_anom}). Compared to the result from \textit{careless}, the anomalous signal in HEWL is slightly lower, but nonetheless of sufficient quality to clearly identify the sulfur and chlorine atoms in the anomalous difference map.

\begin{table}[htbp]
\caption{Sulfur anomalous signal of HEWL from \textit{lawless} and \textit{careless}, at 1.79~\AA.}
\label{tab:hewl_anom}
\begin{threeparttable}
\footnotesize
\begin{tabular}{lcc}
\toprule
 & \textit{lawless} & \textit{careless} \\
\midrule
Mean S peak height ($\sigma$)\tnote{a} & 4.31 & 5.47 \\
Background at C$\alpha$ ($\sigma$)\tnote{b} & 0.95 $\pm$ 0.83 & 1.55 $\pm$ 0.47 \\
S excess over background ($\sigma$) & +3.36 & +3.92 \\
S sites $\geq 3\sigma$ & 10/10 & 10/10 \\
CC($\Delta F$) with S-only model\tnote{c} & 0.158 & 0.086 \\
$R_\mathrm{free}$ & 0.218 & 0.225 \\
Peak at Cl site ($\sigma$)\tnote{d} & 8.2 & 14.0 \\
\bottomrule
\end{tabular}
\begin{tablenotes}
\footnotesize
\item[a] Maximum of the map within 1~\AA\ of each of the 10 S atoms (8 Cys S$\gamma$, 2 Met S$\delta$), averaged.
\item[b] The same measure at C$\alpha$ atoms more than 4~\AA\ from any S; mean $\pm$ s.d.
\item[c] Correlation of observed Bijvoet differences with those calculated with S as the only anomalous scatterer.
\item[d] Cl position of PDB 193L.
\item Model: PDB 193L without waters or ions, refined (\textit{phenix.refine}: rigid body, coordinates, ADPs). \textit{lawless}: data integrated with the beamstop arm masked; rotation scales (spacing 5) with $B$, secondary-beam absorption correction, GPR wavelength normalization, \texttt{REJECT 4 LARGER}. \textit{careless}: poly mode, double-Wilson prior between the Friedel halves, run on the integration without the mask; one overall factor (1.111) puts its two halves on a common scale.
\end{tablenotes}
\end{threeparttable}
\end{table}

\subsection{Better merging statistics do not mean better (neutron) maps}

\begin{figure}[htbp]
    \centering
    \includegraphics[width=1\linewidth]{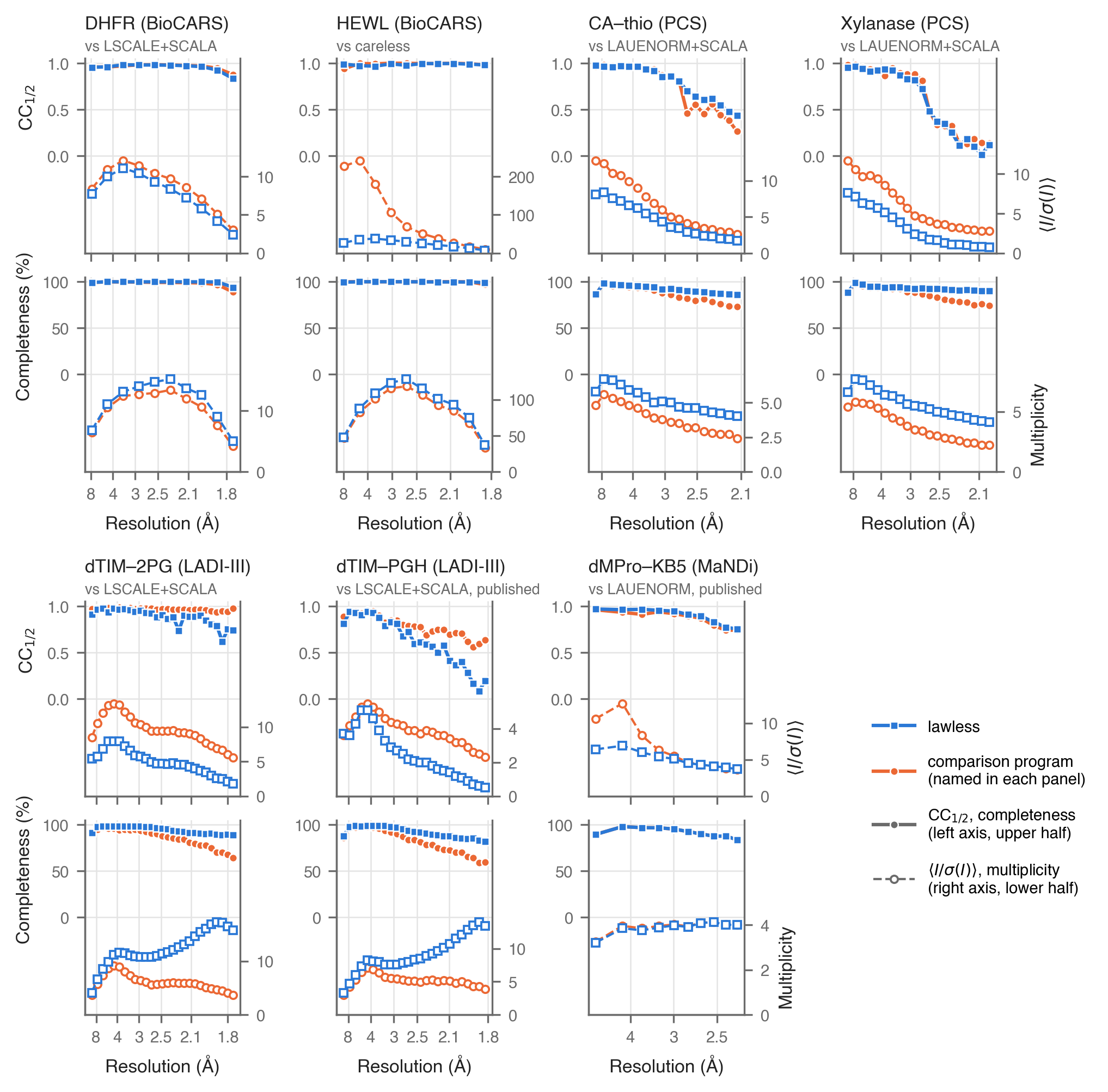}
    \caption{Per-shell statistics for \textit{lawless} (blue squares) and the comparison processing (orange circles).}
    \label{fig:res_shells}
\end{figure}

Table \ref{tab:legacy} and Figure \ref{fig:res_shells} compare merging statistics between published data for the neutron Laue datasets and those of \textit{lawless}. dMPro-KB5 gives quite similar results due to the fact that \textit{LAUENORM} does not reject reflections, and the data processing in \textit{Mantid} at the MaNDi instrument handles outlier rejection on its own, mostly by rejecting reflections that it cannot profile properly\cite{coates_macromolecular_2015}. However, comparison with \textit{LSCALE/SCALA} from the LADI-III datasets is notable, as well as the xylanase dataset from PCS. On paper, \textit{lawless} merging statistics look worse than the published data, and it is clear why: \textit{lawless} keeps 10-30\% more reflections at 2-3x the multiplicity, at the cost of higher $R_{pim}$ and lower $I/\sigma$. In the published work \cite{kelpsas_perdeuteration_2019}, LSCALE/SCALA were optimized to maximize the merging quality and completeness of the datasets (see Table 4 in Kelpsas, et al). In the case of dTIM-PGH, whose data were quite weak, the \textit{lawless} merging statistics look abysmal compared to the previous work. But analysis of the neutron maps tell a different story. 

In neutron MX, poor merging statistics are not the main challenge; \textit{completeness} is. Low neutron flux and long wavelengths relative to X-rays are the primary reasons for this. Longer neutron wavelengths improve scattering probability, but make smaller-radius Ewald spheres, leading to blind spots in reciprocal space that make it difficult to collect complete datasets on low-symmetry crystals. Thus, completeness on such collected datasets tend to be low. But therein lies a conundrum: is it better to throw away "bad", but unique, reflections, or keep them for the sake of completeness? 

\begin{figure}[htbp]
    \centering
    \includegraphics[width=1\linewidth]{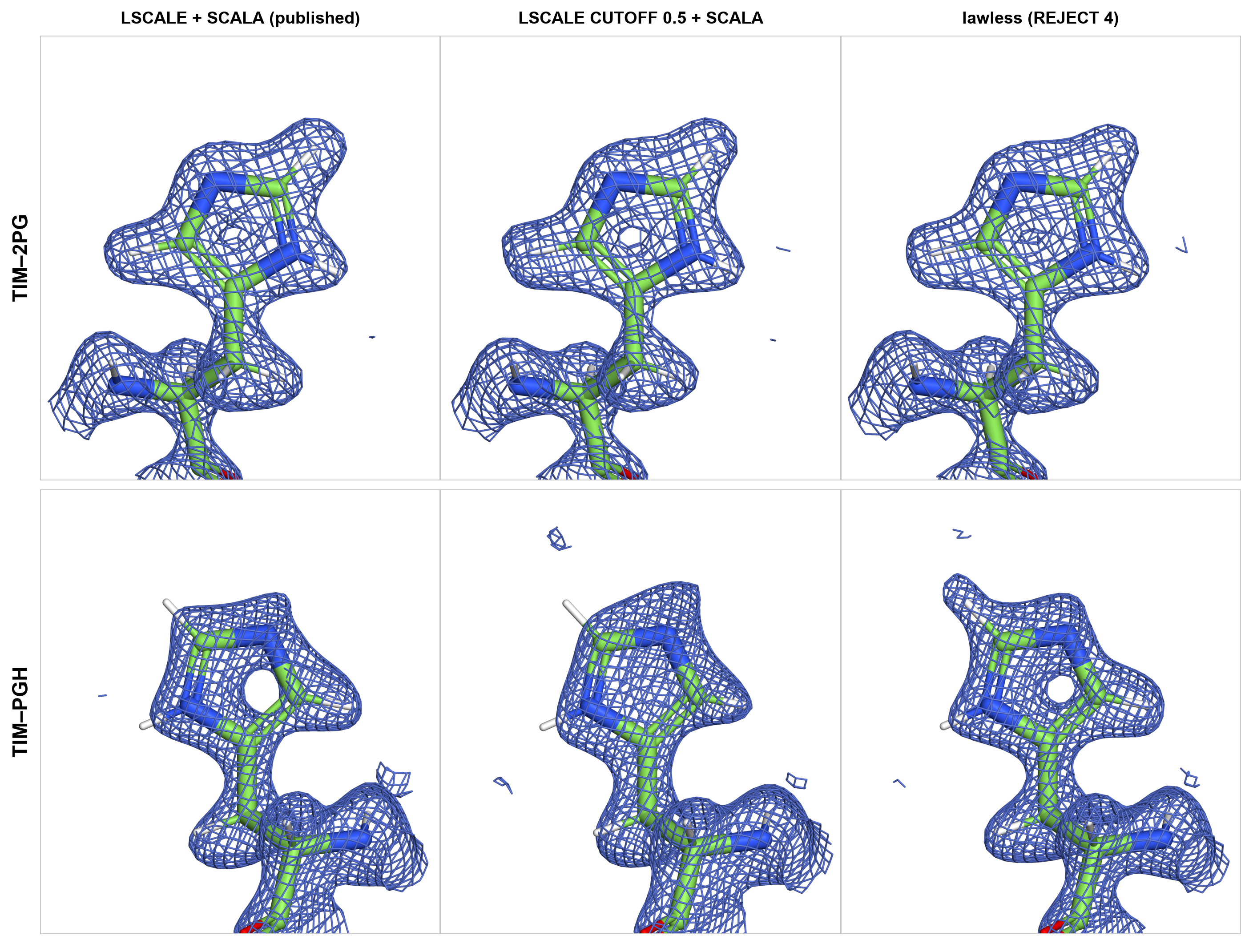}
    \caption{His187 (chain A) of dTIM in the 2-phosphoglycolate (dTIM–2PG, top) and phosphoglycolohydroxamate (dTIM–PGH, bottom) complexes. Neutron $2mFo–DFc$ maps (blue mesh) at $1.5\sigma$ (full-cell $\sigma$), each from its own joint X-ray/neutron refinement, from the published \textit{LSCALE+SCALA} data (left; dTIM–2PG from the published 46-image integration, dTIM–PGH as deposited, \texttt{CUTOFF} 2.0), \textit{LSCALE} with \texttt{CUTOFF} 0.5 + \textit{SCALA} (middle) and \textit{lawless} with \texttt{REJECT} 4 (right).}
    \label{fig:his187}
\end{figure}

\begin{figure}[htbp]
    \centering
    \includegraphics[width=1\linewidth]{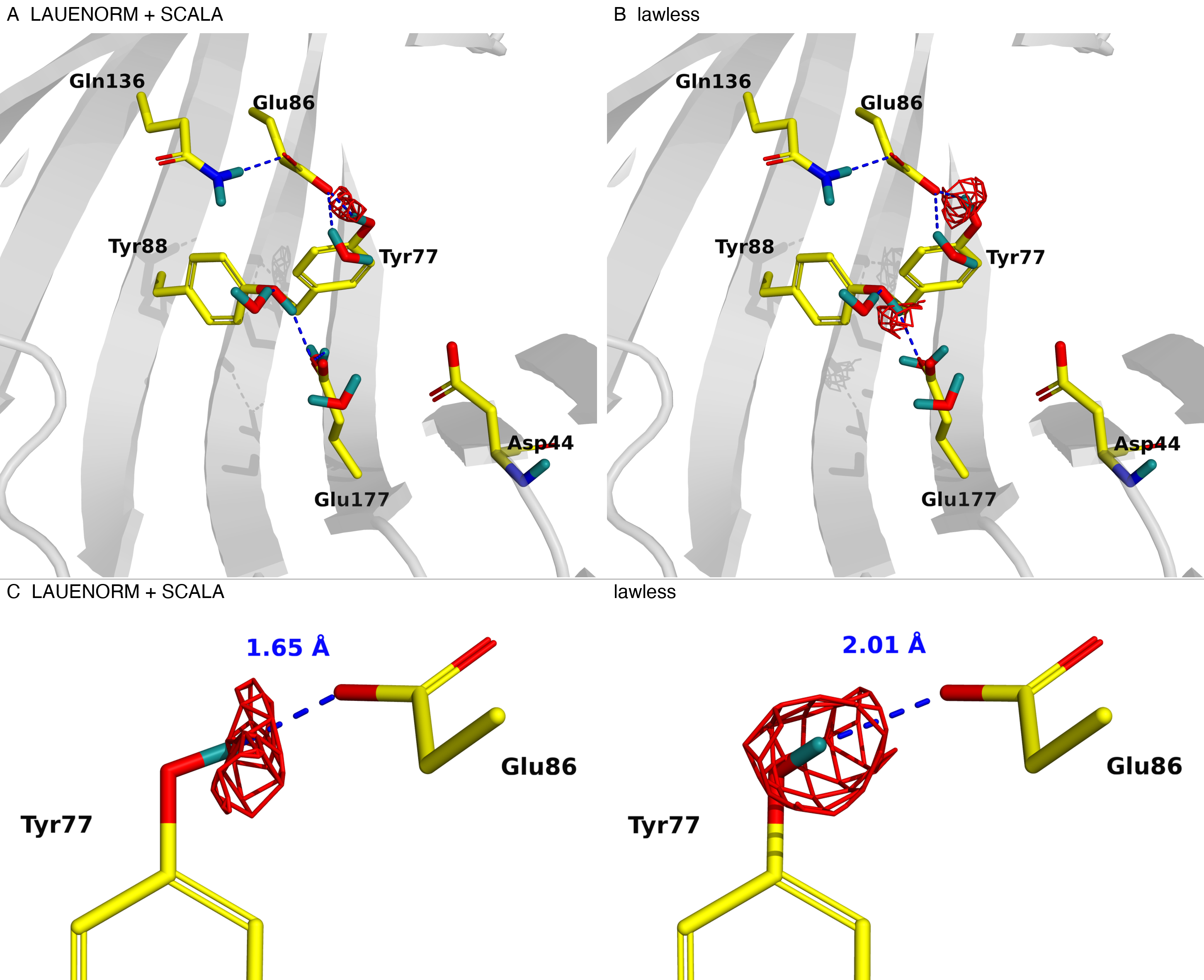}
    \caption{Active site of xylanase N44D (PCS; the scene of Wan et al. 2015, Fig. 1B). Neutron $mFo–DFc$ omit maps (red mesh, 3.0$\sigma$) for the Tyr77 and Tyr88 hydroxyl H/D, from the \textit{LAUENORM+SCALA} data (A) and the \textit{lawless} data (B).}
    \label{fig:xyn_omit}
\end{figure}

An analysis of the maps derived from neutron data is the "truest" measure of data quality, and since neutron MX data is used to find hydrogen/deuterium locations, the focus was put on them. Omit maps were generated on all exchangeable D sites with the \textit{lawless} and \textit{LSCALE} datasets, and the differences in omit peak heights for the D sites were evaluated versus the published data (Table \ref{tab:domit}). \textit{lawless} provided considerable improvements in omit density for all but dMPro, where the improvements were small. The His187 residue of dTIM (Fig. \ref{fig:his187}) is an illustrative example. In the dTIM-PGH map, the C-$\epsilon$1 D-site of His187 has higher prominence in the \textit{lawless} dataset. Likewise, the omit density in the xylanase dataset (Fig. \ref{fig:xyn_omit}) is much more prominent in the \textit{lawless} dataset, such that it pushes the Tyr77 hydroxyl D that interacts with Glu86 to a different orientation. While this has little bearing on the conclusions of the published work, it very well could in other studies.

\begin{table*}[htbp]
\caption{Nuclear density at exchangeable D sites in omit maps.}
\label{tab:domit}
\begin{threeparttable}
\footnotesize
\setlength{\tabcolsep}{2.8pt}
\begin{tabular}{lcrcrccc}
\toprule
 & \multicolumn{2}{c}{Amide D} & \multicolumn{2}{c}{Other exchangeable D} & \multicolumn{3}{c}{Neutron refinement\tnote{a}} \\
\cmidrule(lr){2-3}\cmidrule(lr){4-5}\cmidrule(lr){6-8}
Neutron data & Mean ($\sigma$) & $\Delta$ ($\sigma$) [95\% CI] & Mean ($\sigma$) & $\Delta$ ($\sigma$) [95\% CI] & $d_\mathrm{min}$ (\AA) & $R_\mathrm{work}$ & $R_\mathrm{free}$ \\
\midrule
\multicolumn{8}{l}{\textbf{Xylanase, PCS} } \\
\multicolumn{8}{l}{107 amide D, 115 other exch. D\tnote{b}} \\
\textit{LAUENORM}+\textit{SCALA} & 1.35 & -- & 1.31 & -- & 2.00 & 0.282 & 0.321 \\
\textit{lawless} & 1.94 & +0.59 [0.50, 0.69] & 1.72 & +0.40 [0.30, 0.51] & 2.00 & 0.240 & 0.284 \\
\textit{lawless}, matched\tnote{c} & 1.88 & +0.53 [0.45, 0.62] & 1.68 & +0.36 [0.26, 0.47] & 2.00 & 0.247 & 0.290 \\
\addlinespace
\multicolumn{8}{l}{\textbf{dTIM--2PG, LADI-III}} \\
\multicolumn{8}{l}{210 amide D, 142 other exch. D} \\
\textit{LSCALE}+\textit{SCALA}, published & 3.04 & -- & 1.28 & -- & 1.77 & 0.201 & 0.226 \\
\textit{lawless} & 3.27 & +0.23 [0.14, 0.31] & 1.51 & +0.23 [0.14, 0.32] & 1.77 & 0.199 & 0.221 \\
\textit{lawless}, matched\tnote{c} & 3.21 & +0.16 [0.08, 0.24] & 1.50 & +0.22 [0.14, 0.31] & 1.77 & 0.193 & 0.218 \\
\addlinespace
\multicolumn{8}{l}{\textbf{dTIM--PGH, LADI-III}} \\
\multicolumn{8}{l}{201 amide D, 166 other exch. D} \\
\textit{LSCALE}+\textit{SCALA}, published & 1.50 & -- & 0.29 & -- & 1.80 & 0.257 & 0.299 \\
\textit{lawless} & 1.63 & +0.13 [0.06, 0.21] & 0.39 & +0.10 [0.02, 0.18] & 1.80 & 0.263 & 0.299 \\
\textit{lawless}, matched\tnote{c} & 1.57 & +0.07 [0.01, 0.13] & 0.35 & +0.07 [0.00, 0.13] & 1.80 & 0.253 & 0.287 \\
\addlinespace
\multicolumn{8}{l}{\textbf{dMPro--KB5, MaNDi}} \\
\multicolumn{8}{l}{292 amide D, 237 other exch. D} \\
\textit{LAUENORM}, published & 1.35 & -- & 1.06 & -- & 2.30 & 0.228 & 0.268 \\
\textit{lawless} & 1.38 & +0.03 [$-$0.01, 0.07] & 1.12 & +0.06 [0.03, 0.10] & 2.30 & 0.230 & 0.275 \\
\textit{lawless}, matched\tnote{c} & 1.35 & +0.01 [$-$0.02, 0.03] & 1.13 & +0.07 [0.05, 0.09] & 2.30 & 0.225 & 0.269 \\
\bottomrule
\end{tabular}
\begin{tablenotes}
\footnotesize
\item Neutron omit maps were computed from the X-ray starting model with all H/D atoms removed. Mean: peak height in map r.m.s.\ at the omitted, fully occupied sites. $\Delta$: mean paired per-site difference from published data, with its bootstrap 95\% confidence interval.
\item[a] Joint X-ray/neutron refinement against each row's data (\textit{phenix.refine} \cite{liebschner_improved_2023}).
\item[b] H/D-exchanged data: sites the deposited model (PDB 4XPV) gives a D occupancy $\geq 0.5$. At the 74 amides it gives a D occupancy $<0.5$ (mostly H, negative scattering length) \textit{lawless} is $-0.58\sigma$ [$-0.72$, $-0.43$] lower ($-1.62$ vs $-1.04\sigma$), i.e. the H/D contrast is larger in both directions.
\item[c] \textit{lawless} run restricted to the reflections of the first row.
\end{tablenotes}
\end{threeparttable}
\end{table*}

\subsection{What is \textit{lawless} doing that improves neutron maps?}

It was demonstrated above that \textit{lawless} provides better nuclear density maps than previous data processing routines, even when the merged data numbers look worse. It mostly does this by following the advice that has been advocated by Diederichs et al. for years \cite{diederichs_better_2013, karplus_assessing_2015}: don't throw away data for the sake of better merging statistics. The gentler outlier removal mechanism of \textit{lawless}, compared to \textit{LSCALE}, makes this possible. Table \ref{tab:stats_omit} summarizes how merging statistics and omit map D-site quality are correlated in three datasets. Completeness and multiplicity are the only two merging statistics that positively correlate to improved nuclear density.

\begin{table}[htbp]
\caption{Rank correlation of merging statistics with nuclear density at omitted D sites.}
\label{tab:stats_omit}
\begin{threeparttable}
\footnotesize
\setlength{\tabcolsep}{3pt}
\begin{tabular}{lcccccc}
\toprule
 & \multicolumn{2}{c}{dTIM--PGH} & \multicolumn{2}{c}{Xylanase} & \multicolumn{2}{c}{dTIM--2PG} \\
\cmidrule(lr){2-3}\cmidrule(lr){4-5}\cmidrule(lr){6-7}
Statistic & Amide & Other & Amide & Other & Amide & Other \\
\midrule
Completeness, outer shell & +0.60 & +1.00 & +1.00 & +0.40 & +1.00 & +1.00 \\
Multiplicity, outer shell & +0.71 & +0.83 & +0.20 & +0.80 & +0.50 & +0.50 \\
\addlinespace
$R_\mathrm{pim}$, overall & +0.60 & +1.00 & +0.80 & +0.80 & +1.00 & +1.00 \\
$R_\mathrm{pim}$, outer shell & +0.66 & +0.94 & +0.80 & +0.80 & +1.00 & +1.00 \\
\addlinespace
CC$_{1/2}$, overall & $-$0.77 & $-$0.94 & $-$0.40 & $-$0.60 & $-$1.00 & $-$1.00 \\
CC$_{1/2}$, outer shell & $-$0.83 & $-$0.89 & $-$0.20 & $-$0.80 & $-$1.00 & $-$1.00 \\
\addlinespace
$\langle I/\sigma(I)\rangle$, overall & $-$0.37 & $-$0.94 & $-$0.80 & $-$0.80 & $-$1.00 & $-$1.00 \\
$\langle I/\sigma(I)\rangle$, outer shell & $-$0.43 & $-$0.89 & $-$0.80 & $-$0.80 & $-$1.00 & $-$1.00 \\
\bottomrule
\end{tabular}
\begin{tablenotes}
\footnotesize
\item Spearman rank correlation, within each data set, between a statistic of the common \textit{lawless} merging pass (no outlier rejection, input $\sigma$) and the mean omit-map peak height at amide and other exchangeable D sites (Table~\ref{tab:domit}) over the processing runs of that data set. A larger positive value goes with more D density.
\end{tablenotes}
\end{threeparttable}
\end{table}

But completeness is not the only factor in play. \textit{lawless} also improves the neutron maps for datasets even after completeness is taken into account. An ablation investigation was done for xylanase and dTIM-PGH (Table \ref{tab:ablation}), where various parameters in \textit{lawless} were removed one at a time and the omit maps measured. In both cases, refinement of the error model gave the biggest improvement in map density. GPR vs Chebyshev made no real difference in dTIM-PGH, which is expected because of the smoothness of the LADI-III spectrum, while it made a significant impact on xylanase due to PCS's chopper system, as described above. 

\section{Conclusion}

Here \textit{lawless} is developed, and presented, as a means to an end: a convenient way to improve the quality of Laue MX data by building Laue-specific features on top of the already-popular \textit{AIMLESS} program. In doing so, this will hopefully enable users to generate better neutron maps for analysis with a minimal amount of effort. Future work will be to incorporate these features into the DIALS framework; a description of the Laue-ToF processing pipeline in DIALS will be presented in due course.

\appendix

\section{Gaussian process regression wavelength normalization in \textit{lawless}}\label{GPR_maths}
\subsection{Place in the scale model}
A comprehensive introduction to Gaussian processes (GP) can be found in \cite{rasmussen_gaussian_2008}. Most of the maths described here can be found in greater detail in Rasmussen \& Williams; this section is to show how it is applied in \textit{lawless}.

The GP curve is fitted once, in a pre-pass, before scaling and applied as a fixed factor. Unlike Chebyshev regression in \textit{lawless}, it adds no parameters to the BFGS refinement and has no derivatives. It is normalized at a reference wavelength $w(\lambda_\mathrm{ref})=1$ (\texttt{NORMLAMREF}).

\subsection{Training samples: leave-one-out ratios}
For each unique reflection $h$ with $\geq2$ accepted observations with $I/\sigma\geq-3$ (to keep weak and negative intensities), every observation $l$ gives one sample of the spectral response at $\lambda_\mathrm{hl}$. What is calculated is the ratio of mean of its symmetry mates $l'$, the mates first corrected by the current curve $w^{(k)}$ ($w^{(k)}$ set to identity on the first cycle):

\begin{equation}
    M_\mathrm{hl} = \frac{\sum\nolimits_{l'\ne l}v_{hl'}I_{hl'}/w^{(k)}(\lambda_{hl'})}{\sum\nolimits_{l'\ne l}v_{hl'}},\: v_{hl'}=\frac{w^{(k)}(\lambda_{hl'})^2}{\sigma^2_{e,hl'}},\: \rho_{hl}=\frac{I_{hl}}{M_{hl}}, u_{hl}=\frac{M_{hl}}{\sigma^2_{e,hkl}}
\end{equation}

$$u\rho = \frac{M}{\sigma^2}\cdot\frac{I}{M} = \frac{I}{\sigma^2}$$

so the bin estimate is

$$r_b = \frac{\sum I_{hl}/\sigma^2}{\sum M_{hl}/\sigma^2}$$

$M_{hl}$ is the mean intensity of the symmetry mates. Leaving $l$ out of its own references removes self-bias. It is defined this way such that $I_{hl}\approx M_{hl}w(\lambda_{hl})$. That makes $\rho_{hl}$ a noisy sample of $w$ at that observation's wavelength; this is what the GP is trained on. With $u = M/\sigma^2$ the bin estimator becomes $r_b = \sum I/\sigma^2 \big/ \sum M/\sigma^2$, a ratio of weighted sums (the ratio estimator).

\subsection{Expected $\sigma$-weights}

Under counting statistics $\sigma^2$ grows with $I$, so an observation that came out low by chance gets a small $\sigma$ and a large weight, and the weighted ratio drifts toward a harmonic mean dominated by the weakest observations. To counteract this, the measured variance is split into an intensity-independent part and $b(\lambda)I_\mathrm{pred}$ where $I_\mathrm{pred}$ is the predicted model intensity:
\begin{equation}
    \sigma^2_e = \max(\sigma^2-b(\lambda)I,0) + b(\lambda)\max(I_\mathrm{pred},0), \quad \sigma^2_e \ge 0.01\sigma^2, \quad I_\mathrm{pred}=w(\lambda)M
\end{equation}

$b(\lambda)$ is the slope of the equation $\sigma^2=a+bI$, a least-squares line through the medians of $\sigma^2$ in 10 intensity quantiles, fitted by wavelength in 8 equal-count wavelength bins ($b<0 \rightarrow0)$. With fewer than 400 observations, $b=0$ everywhere and $\sigma$ is used unchanged.

The first raw-cycle ratio measures the response relative to the wavelength mix of the mates. Dividing the mates by the current curve and refitting (3 cycles) converges to the fixed point where the corrected mates are consistent-- the same condition the Chebyshev least-squares process reaches.

\subsection{Binning}

One training point per wavelength bin. Samples are sorted by $\lambda$ and split into equal-count bins (nominal $N/B$ per bin, $B=\texttt{NORMGPRBINS}$, default 50, max $N/20$), a bin also closing when its width in $\ln\lambda$ reaches $4\Delta_x/B$ ($\Delta_x=$span of $\ln\lambda$). This is done because Laue data can have orders of magnitude more data at the spectral peaks than in the tails; equal-length bins give noisy tail points (the source of the oscillation in Fig. \ref{fig:gp_oscillation}), and pure equal-count bins collapse the tail into one wide bin and leave the ends to extrapolation. 

The ratio estimator $r_b$ and its sandwich variance $\mathrm{var}(r_b)$ are defined by:
\begin{equation}
    r_b=\frac{\sum\nolimits_{i\in b}u_i\rho_i}{\sum\nolimits_{i\in b}u_i},\quad \mathrm{var}(r_b)=\frac{\sum\nolimits_{i\in b}u_i^2(\rho_i-r_b)^2}{(\sum\nolimits_{i\in b}u_i)^2}
\end{equation}

The GP target is in log space, like the Chebyshev implementation in \textit{lawless}:

\begin{equation}
    y_b=\ln r_b, \quad s_b^2=\mathrm{var}(r_b)/r_b^2, \quad x_b=\ln\bar{\lambda}_b, \quad \bar{\lambda}_b=\frac{\sum u_i\lambda_i}{\sum u_i}
\end{equation}

with a floor $s_b^2 \ge \max(10^{-6}, 0.01\ \mathrm{median}(s^2))$. Bins with $r_b\le0$ are dropped.

\subsection{Gaussian-process model}
The GP model, in the warped coordinate $x=\ln\lambda$, is defined by:

  \begin{equation}
    y_b = \mu + f(x_b) + \varepsilon_b,\qquad f \sim \mathcal{GP}\big(0, k(x,x')\big),\qquad
    \varepsilon_b \sim \mathcal{N}(0,\, \alpha\, s_b^2).
  \end{equation}

The reason for the warp: a white-beam spectrum can change quickly over a narrow interval at short wavelength and slowly over a wide interval at longer wavelength. In $\ln\lambda$, one stationary length scale fits both: a length scale $\ell$ in $\ln\lambda$ is $\approx  \ell\lambda$ \AA at wavelength $\lambda$. Thus the log warp made the biggest impact at PCS (0.72-6.30 \AA). In narrower-band neutron instruments such as LADI-III and MaNDi, or the pink-beam BioCARS, the rescaling is effectively linear so the warp has no detriment.

There are two kernels $(r=|x-x'|)$ available in \textit{lawless}:

\begin{equation}
    k_{SE}(r)=\sigma_f^2\exp\Big(-\frac{r^2}{2\ell^2}\Big)\:\mathrm{(default)}, \quad
    k_{3/2}(r)=\sigma^2_f\Big(1+\frac{\sqrt{3}r}{\ell}\exp\Big(-\frac{\sqrt{3}r}{\ell}\Big)\ (\texttt{NORMGPRMATERN})
\end{equation}

The mean is constant, fixed at the inverse-variance weighted mean of the targets $\mu=\sum\nolimits_{b}y_bs_b^{-2}/\sum\nolimits_{b}s_b^{-2}$: away from the data the curve reverts to the mean response.

The hyperparameters $\theta=(\ell,\sigma_f,\alpha)$, where $\alpha$ is an inflation of the bin variances, are chosen together by maximizing the log marginal likelihood

\begin{equation}
    \ln p(\mathbf y\mid\theta) = -\tfrac12\,\tilde{\mathbf y}^{\mathsf T}(K + \alpha S)^{-1}\tilde{\mathbf y}
    - \tfrac12 \ln\lvert K + \alpha S\rvert - \tfrac{B}{2}\ln 2\pi,\qquad
    \tilde y_b = y_b - \mu,\ S = \mathrm{diag}(s_b^2).
\end{equation}

The three hyperparameters are chosen by an exhaustive grid search, keeping the combination with the largest $\ln p(\mathbf y\mid\theta)$ (16 × 9 × 8 = 1152 candidates, each one Cholesky factorisation of a $B \times B$ matrix). The length scale takes 16 log-spaced values from $\ell_{\min} = \max(2,\overline{\Delta x}b,\ 0.02,\Delta_x)$, two typical bin spacings, to $\ell{\max} = 1.5,\Delta_x$, where $\Delta_x$ is the span of the training points in $\ln\lambda$ and $\overline{\Delta x}b$ the median spacing of adjacent bins. A user-given length scale (\texttt{NORMGPRLENGTH} $L$, in \AA) replaces the search, converted as $\ell = L/\lambda\mathrm{ref}$ (or $L/\sqrt{\lambda_{\min}\lambda_{\max}}$ without a reference). The signal amplitude takes 9 values $\sigma_f = \sigma_0\,10^{t}$, $t \in [-0.7, 0.7]$, around $\sigma_0^2 = \max\big(\mathrm{var}(\tilde y) - \overline{s^2},\ 10^{-4}\big)$, the spread of the targets minus their mean noise variance. The noise inflation takes 8 values $\alpha = 10^{t}$, $t \in [-0.3, 1.2]$ (0.5 to 16). A warning is printed when $\ell$ or $\alpha$ is chosen at the edge of its range.

The posterior $g(x)$ is given as the Cholesky decomposition of $K+\alpha S$:

\begin{equation}
    g(x) = \mu + \mathbf k_*^{\mathsf T}(K+\alpha S)^{-1}\tilde{\mathbf y},\qquad
    \mathrm{var}\,g(x) = k(x,x) - \mathbf k_*^{\mathsf T}(K+\alpha S)^{-1}\mathbf k_*.
\end{equation}

\subsection{Use of the Normalization Curve}

The normalization curve $w(\lambda)$ is normalized at the reference wavelength 

  \begin{equation}
    w(\lambda) = \exp\!\big(g(\lambda) - g(\lambda_\mathrm{ref})\big) > 0,\quad
    \frac{\sigma(w)}{w} \simeq \mathrm{sd}\,g(\lambda).
  \end{equation}

The curve is stored as a lookup table on $\max(200,4B)$ points uniform in $\lambda$ over $[\lambda_{min},\lambda_{max}]$, linearly interpolated. $w=1$ outside the fitted range. \texttt{NORMGPR 0 0} takes the entire data range. This is applied as a fixed factor in $g_{hl}$ for all later scaling, merging, and output. \texttt{LAMBDAONLY} outputs data corrected by $w$ alone.

\begin{table*}
\caption{Thaumatin (monochromatic rotation data): \textit{lawless} 0.1.6 and \textit{aimless} 0.8.2 (CCP4~9) run with default settings on the same \textit{XDS}/\textit{POINTLESS} unmerged data.}
\label{tab:thaumatin}
\begin{threeparttable}
\footnotesize
\setlength{\tabcolsep}{1.7pt}
\begin{tabular}{lcc}
\toprule
 & \textit{lawless} 0.1.6 & \textit{aimless} 0.8.2 \\
\midrule
\multicolumn{3}{l}{\textit{Data collection}} \\
Protein & Thaumatin & Thaumatin \\
Instrument & \begin{tabular}[t]{@{}c@{}}BioMAX,\\MAX IV\end{tabular} & \begin{tabular}[t]{@{}c@{}}BioMAX,\\MAX IV\end{tabular} \\
Probe & X-ray & X-ray \\
Method & Rotation & Rotation \\
Wavelength (\AA) & 0.7293 & 0.7293 \\
Images (batches) & 3600 & 3600 \\
Space group & $P4_12_12$ & $P4_12_12$ \\
$a$, $b$, $c$ (\AA) & 57.9,\,57.9,\,150.3 & 57.9,\,57.9,\,150.3 \\
$\alpha$, $\beta$, $\gamma$ ($^\circ$) & 90,\,90.0,\,90 & 90,\,90.0,\,90 \\
Data Reduction & \textit{XDS} & \textit{XDS} \\
\addlinespace
\multicolumn{3}{l}{\textit{Scaling and merging}} \\
$\lambda$ normalization & -- & -- \\
Resolution (\AA) & 45.8--1.45 & 45.8--1.45 \\
Outer shell (\AA) & 1.48--1.45 & 1.48--1.45 \\
Observations & 911\,096 & 911\,096 \\
Unique reflections & 43\,278 & 43\,278 \\
Multiplicity & 21.1 (2.5) & 21.1 (2.5) \\
Completeness (\%) & 93.8 (50.3) & 93.8 (50.3) \\
$\langle I/\sigma(I)\rangle$ & 50.2 (4.5) & 50.2 (4.5) \\
CC$_{1/2}$ & 1.000 (0.962) & 1.000 (0.962) \\
$R_\mathrm{meas}$ & 0.040 (0.189) & 0.040 (0.189) \\
$R_\mathrm{pim}$ & 0.008 (0.104) & 0.008 (0.104) \\
$\langle\chi^2\rangle$ & 0.94 (0.66) & 0.94 (0.66) \\
$d$, CC$_{1/2}{=}0.3$ (\AA) & 1.45 & 1.45 \\
SdFac, SdB, SdAdd & 0.96,\,0.09,\,-0.01 & 0.96,\,0.09,\,-0.01 \\
Outliers rejected & 14 & 14 \\
\bottomrule
\end{tabular}
\begin{tablenotes}
\footnotesize
\item Values in parentheses are for the outermost resolution shell.
\item The merged intensities of the two programs agree to within $1.5\times10^{-5}\,\sigma$ (Section~5.1).
\end{tablenotes}
\end{threeparttable}
\end{table*}

\begin{table*}
\caption{One-factor ablation of the \textit{lawless} deck: effect on nuclear density at omitted D sites.}
\label{tab:ablation}
\begin{threeparttable}
\footnotesize
\setlength{\tabcolsep}{3.5pt}
\begin{tabular}{lccccr}
\toprule
Variant & Amide D ($\sigma$) & Other exch.\ D ($\sigma$) & CC$_{1/2}$ & $R_\mathrm{free}$ & Outliers \\
\midrule
\multicolumn{6}{l}{\textbf{Xylanase (PCS, 2.0~\AA)\tnote{b}}} \\
Final deck\tnote{c} & 1.94 & 1.72 & 0.806 & 0.3229 & 370 \\
SD model fixed\tnote{d} & $-$0.41 [$-$0.54, $-$0.28] & $-$0.44 [$-$0.55, $-$0.33] & 0.882 & 0.3863 & 17\,798 \\
Chebyshev normalization\tnote{a} & $-$0.33 [$-$0.41, $-$0.25] & $-$0.35 [$-$0.44, $-$0.26] & 0.795 & 0.3526 & 3\,646 \\
\texttt{REJECT 2} & $-$0.11 [$-$0.16, $-$0.05] & $-$0.11 [$-$0.15, $-$0.07] & 0.850 & 0.3271 & 6\,291 \\
\texttt{REJECT 3} & $-$0.03 [$-$0.05, $-$0.00] & $-$0.01 [$-$0.04, 0.02] & 0.820 & 0.3234 & 1\,297 \\
\texttt{REJECT NONE} & $-$0.01 [$-$0.03, 0.02] & $-$0.01 [$-$0.03, 0.02] & 0.802 & 0.3231 & 0 \\
Negatives set to 0\tnote{e} & $-$0.06 [$-$0.09, $-$0.03] & $-$0.04 [$-$0.07, $-$0.01] & 0.867 & 0.3249 & 455 \\
\addlinespace
\multicolumn{6}{l}{\textbf{dTIM--PGH (LADI-III, 1.80~\AA)}} \\
Final deck\tnote{c} & 1.63 & 0.39 & 0.826 & 0.2755 & 512 \\
SD model fixed\tnote{d} & $-$0.23 [$-$0.29, $-$0.16] & $-$0.08 [$-$0.15, $-$0.01] & 0.795 & 0.3049 & 9\,072 \\
Chebyshev normalization\tnote{a} & +0.01 [$-$0.01, 0.03] & +0.02 [$-$0.01, 0.04] & 0.808 & 0.2714 & 785 \\
\texttt{REJECT 2} & $-$0.16 [$-$0.21, $-$0.11] & $-$0.02 [$-$0.07, 0.03] & 0.864 & 0.2886 & 12\,905 \\
\texttt{REJECT 3} & $-$0.03 [$-$0.05, $-$0.01] & +0.01 [$-$0.02, 0.03] & 0.841 & 0.2797 & 2\,068 \\
\texttt{REJECT NONE} & +0.01 [$-$0.00, 0.02] & $-$0.01 [$-$0.03, 0.00] & 0.821 & 0.2758 & 0 \\
Negatives set to 0\tnote{e} & $-$0.08 [$-$0.11, $-$0.04] & $-$0.06 [$-$0.09, $-$0.02] & 0.879 & 0.2856 & 449 \\
\bottomrule
\end{tabular}
\begin{tablenotes}
\footnotesize
\item Each row changes one setting of the final deck; the omit map is computed from the data set’s common D-omitted start model. Final deck: mean omit-map peak height (map r.m.s.) at 107 amide and 115 other exchangeable D sites (xylanase: sites with deposited D occupancy $\geq 0.5$, PDB 4XPV) and 201 / 166 (dTIM--PGH, fully occupied sites). Other rows: mean paired per-site difference from the final deck, with its bootstrap 95\% confidence interval. CC$_{1/2}$ and outliers from each run's own \textit{lawless} log; $R_\mathrm{free}$: the common start model against each data set (omit map, no refinement). The established processing gives 1.35 / 1.31 (\textit{LAUENORM}+\textit{SCALA}, xylanase) and 1.50 / 0.29 (\textit{LSCALE}+\textit{SCALA}, published, dTIM--PGH).
\item[a] As in the main-text comparison of GPR and Chebyshev normalization: two ranges of degree 6 split at 1.5~\AA\ (xylanase), degree 9 (dTIM--PGH).
\item[b] At the 74 amides with deposited D occupancy $<0.5$ (mostly H) the final deck gives $-$1.62$\sigma$; the change (positive = less H/D contrast): SD model fixed +0.36, Chebyshev +0.20, \texttt{REJECT 2} +0.03, negatives set to 0 +0.10.
\item[c] \texttt{REJECT 4}, GPR wavelength normalization, refined error model (\texttt{SDCORRECTION REFINE}).
\item[d] \texttt{SDCORRECTION NOREFINE 1 0 0}: input $\sigma$ unchanged.
\item[e] Every negative input intensity set to 0 (xylanase: the \texttt{.afout} files \textit{LAUENORM} read).
\end{tablenotes}
\end{threeparttable}
\end{table*}

\FloatBarrier

\begin{acknowledgements}
The greatest of thanks first to Phil Evans, the author of \textit{AIMLESS}, for many helpful discussions. The author would also like to thank Zo{\"e} Fisher for providing the PCS-LANCSE data, Esko Oksanen for providing the dTIM-2PG and dTIM-PGH data, and Andrey Kovalevsky for providing the dMPro-KB5 data; and to Kevin Dalton and Rick Hewitt for helpful discussions about \textit{careless} and the BioCARS datasets.

\end{acknowledgements}

\begin{funding}
List funding organizations, recipients, grant numbers, etc.
\end{funding}

\ConflictsOfInterest{The author declares no conflicts of interest.
}

\DataAvailability{The source code archive for \textit{lawless}, as well as the repository link, can be found at https://doi.org/10.5281/zenodo.23929955. The datasets for HEWL (https://data.sbgrid.org/dataset/1118/) and DHFR (https://data.sbgrid.org/dataset/1117/) can be found online. 
}

\bibliography{iucr} % basename of .bib file

\end{document}